\documentclass[%
 reprint,
 superscriptaddress,
 showpacs,preprintnumbers,
 amsmath,amssymb,
 aps,
 prb,
]{revtex4-1}

\usepackage{graphicx}% Include figure files
\usepackage{dcolumn}% Align table columns on decimal point
\usepackage{bm}% bold math
\usepackage{upgreek}
\usepackage{amsmath}

\begin{document}

%\preprint{APS/123-QED}

\title{Tuning nonlinear second-harmonic generation in AlGaAs nanoantennas via chalcogenide phase change material}% Force line breaks with \\

\author{Tingting Liu}
\email{ttliu@usst.edu.cn}
\affiliation{Institute of Photonic Chips, University of Shanghai for Science and Technology, Shanghai 200093, China}
\affiliation{Centre for Artificial-Intelligence Nanophotonics, School of Optical-Electrical and Computer Engineering, University of Shanghai for Science and Technology, Shanghai 200093, China}

\author{Xinyuan Fang}
\email{xinyuan.fang@usst.edu.cn}
\affiliation{Institute of Photonic Chips, University of Shanghai for Science and Technology, Shanghai 200093, China}
\affiliation{Centre for Artificial-Intelligence Nanophotonics, School of Optical-Electrical and Computer Engineering, University of Shanghai for Science and Technology, Shanghai 200093, China}

\author{Shuyuan Xiao}
\email{syxiao@ncu.edu.cn}
\affiliation{Institute for Advanced Study, Nanchang University, Nanchang 330031, China}
\affiliation{Jiangxi Key Laboratory for Microscale Interdisciplinary Study, Nanchang University, Nanchang 330031, China}

\begin{abstract}
	
The ability to engineer nonlinear optical processes in all-dielectric nanostructures is both of fundamental interest and highly desirable for high-performance, robust, and miniaturized nonlinear optical devices. Herein, we propose a novel paradigm for the efficient tuning of second-harmonic generation (SHG) process in dielectric nanoantennas by integrating with chalcogenide phase change material. In a design with Ge$_{2}$Sb$_{2}$Te$_{5}$ (GST) film sandwiched between the AlGaAs nanoantennas and AlO$_{x}$ substrate, the nonlinear SHG signal from the AlGaAs nanoantennas can be boosted via the resonantly localized field induced by the optically-induced Mie-type resonances, and further modulated by exploiting the GST amorphous-to-crystalline phase change in a non-volatile, multi-level manner. The tuning strategy originates from the modulation of resonant conditions by changes in the refractive index of GST. With a thorough examination of tuning performances for different nanoantenna radii, a maximum modulation depth as high as 540$\%$ is numerically demonstrated. This work not only reveals out the potential of GST in optical nonlinearity control, but also provides promising strategy in smart designing tunable and reconfigurable nonlinear optical devices, e.g., light emitters, modulators, and sensors.

\end{abstract}

%\pacs{42.70.-a, 42.79.-e, 78.67.Pt}% PACS, the Physics and Astronomy
\maketitle

%\tableofcontents

\section{\label{sec1}Introduction}

The recently developed nonlinear nanophotonics provides an attractive platform for the manipulation of strong light-matter interaction at the subwavelength scale, which alleviates phase matching constraints typically of the bulk and reaches remarkable efficiency by the resonantly localized field. It facilitates various nonlinear optical processes such as frequency conversion, wave mixing, and all-optical switching, with many potential applications for coherent light sources, holography, and other ultrafast miniature devices\cite{Kauranen2012, Smirnova2016, Li2017}. Compared with the metallic plasmonic counterparts, the emerging high-index all-dielectric nanostructures have negligible Ohmic losses to withstand much higher pump field intensities, and their fabrications exhibit complete compatibility with the low-cost complementary metal-oxide semiconductor (CMOS) technology\cite{Genevet2017, Baranov2017, Wang2020, Odit2021, Tuz2021}. Moreover, they offer intriguing capabilities to support both the optically-induced electric and magnetic Mie-type resonances that can be utilized to enhance and control the nonlinear phenomena via multipolar interference effects\cite{Kruk2017, Sain2019, Koshelev2020}. Indeed, recent years have witnessed the giant improvement of nonlinear optical processes in all-dielectric nanostructures, showing record high conversion efficiency especially for second-harmonic generation (SHG)\cite{Camacho-Morales2016, Vabishchevich2018, Carletti2018, Frizyuk2019, Koshelev2020a, Li2020, Volkovskaya2020, Anthur2020, Wang2021} and third-harmonic generation (THG)\cite{Yang2015, Grinblat2016, Tong2016, Xu2018, Gao2018, Ban2019, Semmlinger2019, Gandolfi2021}.

The research agenda is now shifting towards achieving tunable and reconfigurable nonlinear optical processes to develop multi-functional devices\cite{Zheludev2012, Minovich2015, Xiao2020}. With an additional degree of freedom compared with the linear optical processes, the complex nonlinearity in all-dielectric nanostructures makes its modulation even more elusive. One of the most intuitive solution is to engineer the geometric parameters of designed nanostructure, the polarization and intensity of incident light, which has been demonstrated to shape the nonlinear wavefront especially the radiation pattern of harmonic generation in previous works\cite{Carletti2016, Wang2017, Melik-Gaykazyan2018, Rocco2018, Xu2019, Xu2020, Frizyuk2021}. On the other hand, motivated by the recent progress of active nanophotonics, a great interest has been triggered into the study of dynamical control of harmonic generation in dielectric nanostructures using external stimuli\cite{Che2020}. For instance, the modulation of the SHG conversion efficiency up to $60\%$ has been experimentally demonstrated in AlGaAs nanoantennas by thermally tuning the refractive index, and further extended to photo-induced permittivity changes featured with ultrafast response\cite{Celebrano2021, Pogna2021}. Another example has utilized a liquid crystal matrix as the upper media of AlGaAs metasurfaces, and the SHG signal can be activated on or turned off by electrically switching the liquid crystal orientation\cite{Rocco2020}. However, most of the above tuning strategies are volatile and can not maintain stable after removing the external stimuli, while the non-volatility is in increasing demand for up-to-date tunable and reconfigurable optical devices.

In this work, we propose a novel paradigm for tuning the nonlinear SHG in AlGaAs nanoantennas by introducing the chalcogenide phase change material Ge$_{2}$Sb$_{2}$Te$_{5}$ (GST) film. GST is featured with the high refractive index contrast between its amorphous and crystalline phases across the visible and infrared spectrum, and in sharp contrast to the above-mentioned tuning strategies such as thermal tuning or utilizing liquid crystals, it affords an entirely different, non-volatile approach where the induced change maintains stable even after the removal of external stimuli\cite{Wuttig2017, Ding2019}. By continuously exploiting the intermediate phases of GST in the phase change process, we show that the progressive changes in optical properties of GST lead to the gradual variation in the resonant properties of the AlGaAs nanoantennas in both the linear and nonlinear regions. With a thorough examination of tuning performances for different nanoantenna radii, a maximum modulation depth of the nonlinear SHG efficiency up to 540$\%$ is numerically demonstrated in the AlGaAs nanoantennas during the full process of GST amorphous-to-crystalline phase change. This work shows great prospects in designing tunable and reconfigurable nonlinear devices for a broad class of optical signal processing.
 
\section{\label{sec2}Model and Method}

\begin{figure*}[htbp]
\centering
\includegraphics% Here is how to import EPS art
[scale=0.20]{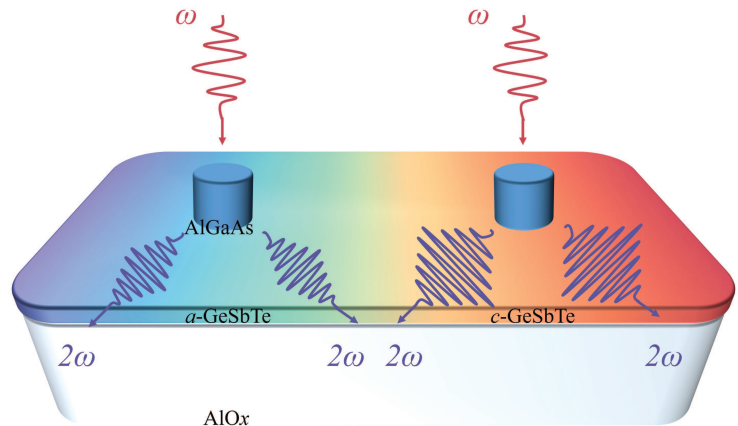}
\caption{\label{fig1} Conceptual schematic of the tunable nonlinear dielectric nanostructure composed of AlGaAs nanoantennas, GST film, and AlO$_{x}$ substrate, where the full process of GST amorphous-to-crystalline phase change can be exploited for tuning the SHG process in AlGaAs nanoantennas in a non‐volatile, multi-level way.}
\end{figure*}

The basic concept is schematically illustrated in Fig. \ref{fig1}. The nonlinear optical process occurs in the AlGaAs nanoantennas: when irradiated with intense laser light at fundamental wavelength, the optically-induced electric and magnetic Mie-type resonances boost the SHG emission from the AlGaAs nanoantennas via the resonantly localized field. To modulate the SHG signal, a GST film is inserted between the AlGaAs nanoantennas and AlO$_{x}$ substrate: under the external stimuli, the active material GST undergoes a non‐volatile phase change from amorphous (a-GST) to crystalline state (c-GST), and its refractive index drastically changes during the process in a broad range of wavelengths. Due to the strong dependence of the electric and magnetic resonances on the surrounding dielectric environment, the tunable optical response to material refractive index changes can be expected in both the linear and nonlinear regions.  

\begin{figure*}[htbp]
\centering
\includegraphics% Here is how to import EPS art
[scale=0.48]{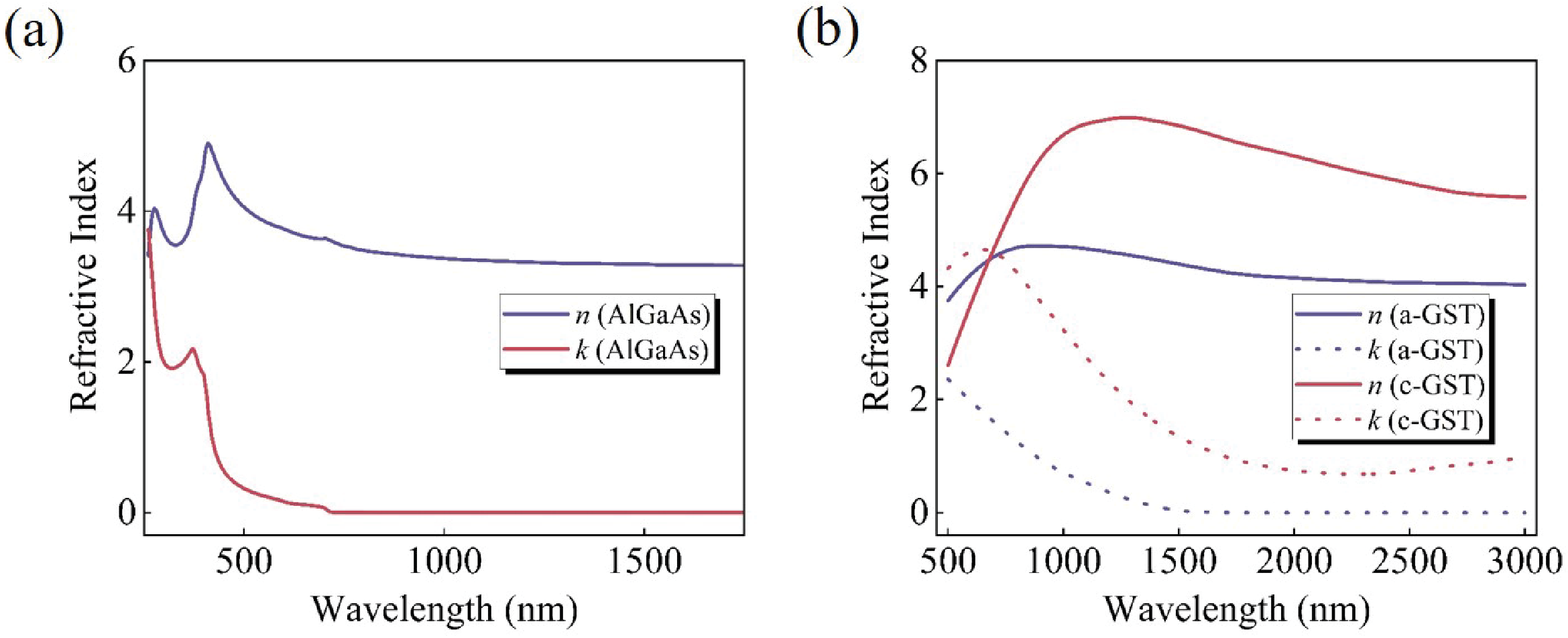}
\caption{\label{fig2} (a) The refractive index of AlGaAs. (b) The refractive indices of GST for its amorphous phase (blue line) and crystalline phase (red line), respectively. The optical constants are extracted from the experimental measurement using spectroscopic ellipsometry\cite{Papatryfonos2021, Cao2019}.}
\end{figure*}

To investigate the optical responses of the nonlinear dielectric nanostructure, we perform the three-dimensional electromagnetic simulations with the finite element method solver of COMSOL Multiphysics. In the numerical model, the isolated AlGaAs nanoantennas shaped as nanocylinders of 300 nm height and different radii are equally spaced by 3 $\upmu$m and placed on a 50-nm-thick GST film, supported by the semi-infinite AlO$_{x}$ substrate. A $x$-polarized plane wave with 1550 nm central wavelength is incident from the air upper half-space, and the perfectly matched layers are imposed in both the vertical (along the $z$-axis) and lateral directions (along the $x$- and $y$-axis). In this design, the semiconductor AlGaAs has been recently investigated by several groups since it possesses high transparency in a broad spectral window up to the far infrared and giant second-order nonlinear susceptibility\cite{Carletti2015, Gili2016}. For the dispersion of AlGaAs, here we adopt the wavelength-dependent complex-value refractive index which is consistent with the experimental measurement\cite{Papatryfonos2021}, as plotted in Fig. \ref{fig2}(a). Due to the zinc blende crystalline structure, AlGaAs has the only non-vanishing terms of the second-order nonlinear susceptibility tensor $\chi_{ijk}^{(2)}$ with $i\neq j\neq k$, which is taken to be 100 pm/V\cite{Ohashi1993}. The nonlinear SHG from the AlGaAs nanoantennas is thus calculated by two steps in the undepleted pump approximation: in the first step, we calculate the linear optical response at the fundamental wavelength, in the second step, the second-order nonlinear polarizabilities, induced by the electric field at the fundamental wavelength using the above-mentioned $\chi^{(2)}$ tensor is then employed as a source to obtain the generated SHG field,
\begin{equation}
P_{i}^{(2\omega)}=\varepsilon_{0}\chi_{ijk}^{(2)}E_{j}^{(\omega)}E_{k}^{(\omega)}\,, i\neq j\neq k.\label{eq1}
\end{equation}
GST is an active material with a crystallization temperature $T_{\text{c}}$ of 433 K and a melting temperature $T_{\text{m}}$ of 873 K, while the phase change in both directions can be realized within only about few tenths of nanoseconds by using external thermal stimuli such as heat, electrical, or optical pulses. For its amorphous and crystalline phases, the complex-value refractive indices in Fig. \ref{fig2}(b) are extracted from the experimental data\cite{Cao2019}. When the partial crystallization of GST is taken into consideration, i.e., by varying the the thermal annealing time or the optical pulse intensity applied to the GST film, the effective permittivity of such phase change material is gradually changed owing to the formation of nucleation in a-GST, and therefore a continuous multi-level control of the nonlinear SHG in the AlGaAs nanoantennas can be realized by exploiting these intermediate phases. The effective permittivity at different crystallization ratios can be modeled by the Lorentz-Lorenz relation\cite{Tian2019, Zhu2020, Zhou2020},
\begin{equation}
\frac{\varepsilon_{\text{eff}}-1}{\varepsilon_{\text{eff}}+2}
=m\times\frac{\varepsilon_{\text{c-GST}}-1}{\varepsilon_{\text{c-GST}}+2}
+(1-m)\times \frac{\varepsilon_{\text{a-GST}}-1}{\varepsilon_{\text{a-GST}}+2},\label{eq2}
\end{equation}
where $\varepsilon_{\text{a-GST}}$ and $\varepsilon_{\text{c-GST}}$ are the complex-value permittivity of a-GST and c-GST, respectively, which can be calculated with the refractive indices in Fig. \ref{fig2}(b), and $m$ is the crystallization ratio of GST, ranging from 0 to 1 for the intermediate phases. 

It is noteworthy that only the semiconductor AlGaAs is considered as the harmonic source in the proposed design, because AlO$_{x}$ has negligible $\chi^{(2)}$ and SHG signal could not be found in the amorphous and nonordered crystalline films of GST except in a very special case of the oriented GST grains induced by polarized laser pulses\cite{Meng2020}. The AlGaAs-on-AlO$_{x}$ nanoantennas can be grown by molecular beam epitaxy on a GaAs wafer with subsequent electron beam lithography and dry etching. Then a peel-off process is applied to the top AlGaAs nanoantennas, and a 50-nm-thick GST film is deposited on top of the AlO$_{x}$ substrate via an unbalanced magnetron sputtering. Finally, the AlGaAs nanoantennas are transferred back and bonded to the GST film. This nanofabrication method has been demonstrated in previous works\cite{Camacho-Morales2016}, and all the above realistically reflects the experimental conditions.

\section{\label{sec3}Results and Discussion}

\begin{figure}[htbp]
\centering
\includegraphics% Here is how to import EPS art
[scale=0.48]{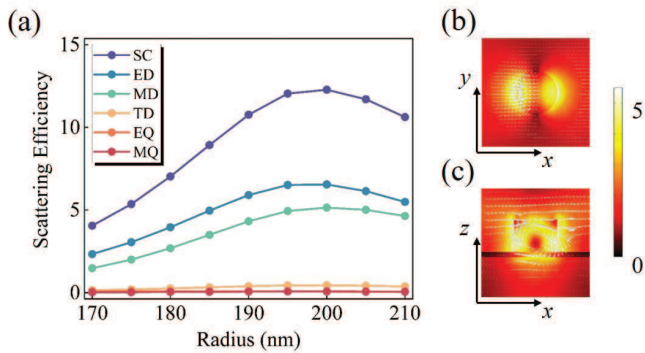}
\caption{\label{fig3} (a) The scattering efficiency spectrum as a function of the nanoantenna radius at the fundamental wavelength of 1550 nm, with the multipolar expansion terms of the linear optical response in Cartesian coordinates. The denotations of SC, ED, MD, TD, EQ, MQ represent total scattering efficiency, the contributions of electric dipole, magnetic dipole, toroidal dipole, electric quadrupole, and magnetic quadrupole, respectively. Electric field distributions (b) in the $x$-$y$ plane at half of the nanoantenna height and (c) in the $x$-$z$ plane for the radius $r=200$ nm, overlaid with the electric field direction vector.}
\end{figure}

For a specific nonlinear nanostructure, the conversion efficiency of the harmonic generation process is primarily determined by the resonantly localized field at the fundamental wavelength. Guided by this principle, we firstly investigate the linear scattering of AlGaAs nanoantennas at the optical communication wavelength of 1550 nm. As is well-known, the position of the Mie-type resonances is highly sensitive to the aspect ratio of the nanocylinder, therefore we sweep the radius from 170 nm to 210 nm for the fixed height of 300 nm. The GST film is initially set to its amorphous phase. Fig. \ref{fig3}(a) shows the simulated scattering efficiency spectrum as a function of the nanoantenna radius, where the scattering efficiency is defined as the scattering cross section normalized by the geometric area of the nanocylinder, 
\begin{equation}
\eta_{\text{SCAT}}=\frac{\int_{A}\vec{S}_{\text{SCAT}}\cdot\hat{n}\, da}{I_{0}\times\pi r^{2}},\label{eq3}
\end{equation}
where $\vec{S}_{\text{SCAT}}$ is the Poynting vector of the fundamental field, $\hat{n}$ is the unit vector normal to a surface $A$ enclosing the nanoantennas, and $I_{0}$ is the intensity of incident light with 1 GW/cm$^{2}$ adopted in the calculations. To further characterize the linear scattering by multipolar modes excited in the nanoantennas, we perform the multipolar expansion in the Cartesian coordinates using the displacement currents induced by the incident electric field. It can be observed that the maximum scattering efficiency occurs for the radius around $r=200$ nm, which results from the leading contributions of electric dipole and magnetic dipole. Figs. \ref{fig3}(b) and \ref{fig3}(c) show the electric field overlaid with its direction vector in the $x$-$y$ and $x$-$z$ planes, respectively. The magnetic dipole character of the resonance can be easily identified by the electric field loop in the vertical direction, and the slight leak into the GST film and the AlO$_{x}$ substrate increase the contribution of electric dipole compared to that in the suspended nanoantennas\cite{Carletti2015, Gili2016}. These two electric and magnetic Mie-type resonances and the associated resonantly localized field provide the platform for enhancing the SHG emission from the AlGaAs nanoantennas.

\begin{figure}[htbp]
\centering
\includegraphics% Here is how to import EPS art
[scale=0.48]{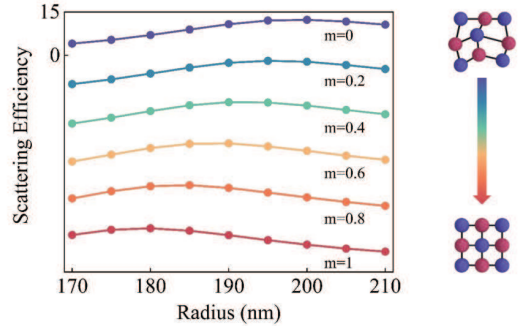}
\caption{\label{fig4} The scattering efficiency spectra as a function of the nanoantenna radius at the fundamental wavelength of 1550 nm during the full process of GST amorphous-to-crystalline phase change. From the top to the bottom are different crystallization ratios, $m=0$ (a-GST), $m=0.2$, 0.4, 0.6, 0.8 (partial crystalline), and $m=1$ (c-GST), respectively.}
\end{figure}

With the aim of achieving tunability and reconfigurability, we explore the dependence of the above linear scattering on the phase change of GST in a step by step manner considering the partial crystallization. When the embedded GST film is in different crystallization conditions, Fig. \ref{fig4} shows the simulated scattering efficiency spectra as a function of the nanoantenna radius at the fundamental wavelength of 1550 nm. As the crystallization ratio of GST increases from $m=0$ (a-GST) to $m=1$ (c-GST), the progressive changes in optical properties of GST lead to the gradual variation in the scattering efficiency of the AlGaAs nanoantennas for every different radii. For radii $r=200$ and 210 nm, the scattering efficiency $\eta_{\text{SCAT}}$ shows a general downward trend, while for other radii $r=170$ nm, 180 nm, and 190 nm, $\eta_{\text{SCAT}}$ displays an overall upward trend during the full process of GST amorphous-to-crystalline phase change. It can be obviously observed that within the range we considered, the radius corresponding to the maximum scattering efficiency exhibits a shift from $r=200$ nm with a-GST to $r=180$ nm with c-GST. As shown in Fig. \ref{fig2}(b), the real part of the complex refractive indices of GST around 1550 nm shows an apparent increase in the amorphous-to-crystalline phase change process, and consequently results in an increase in the effective refractive index of the whole nanostructure, which needs to be compensated by using a relatively small geometry (in the form of the smaller radius here) for maintaining the strong resonance at the fundamental wavelength.

\begin{figure}[htbp]
\centering
\includegraphics% Here is how to import EPS art
[scale=0.48]{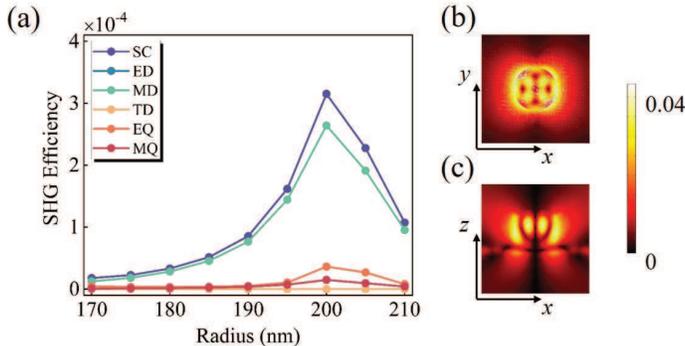}
\caption{\label{fig5} (a) The SHG efficiency spectrum as a function of the nanoantenna radius at the second-harmonic wavelength of 775 nm, with the multipolar expansion terms of the nonlinear optical response in Cartesian coordinates. The denotations of SC, ED, MD, TD, EQ, MQ represent total SHG efficiency, the contributions of electric dipole, magnetic dipole, toroidal dipole, electric quadrupole, and magnetic quadrupole, respectively. Electric field distributions (b) in the $x$-$y$ plane at half of the nanoantenna height and (c) in the $x$-$z$ plane for the radius $r=200$ nm, overlaid with the electric field direction vector.}
\end{figure}

In the following, we extend simulations to the nonlinear region and consider the SHG process in the proposed nanostructure. As we have mentioned above, the nonlinear optical response of the nanoantennas at the second-harmonic wavelength is calculated using the nonlinear polarizabilities generated by the electric field at the fundamental wavelength as a source. The conversion efficiency of the SHG process is calculated as
\begin{equation}
\eta_{\text{SHG}}=\frac{\int_{A}\vec{S}_{\text{SH}}\cdot\hat{n}\, da}{I_{0}\times\pi r^{2}},\label{eq4}
\end{equation}
where $\vec{S}_{\text{SH}}$ is the Poynting vector of the second-harmonic field, and other variables are consistent with those in the linear scattering. The simulated SHG efficiency spectrum in Fig. \ref{fig5}(a) indicates the dependence of the nonlinear conversion efficiency on the nanoantenna radius at half of the fundamental wavelength, i.e. 775 nm. When the radius increases from 170 nm to 210 nm, the SHG efficiency spectrum shows a variation tendency resembling the linear scattering efficiency. Compared with the broad scattering efficiency spectrum in Fig. \ref{fig3}(a), the SHG efficiency shows a relatively sharp peak due to the second-harmonic nature. The SHG efficiency reaches its peak value of $\sim3\times10^{-4}$ for the radius $r=200$ nm which is the same radius corresponding to the maximum linear scattering efficiency. Such good correspondence between the nonlinear and linear optical processes stems from the fact that it is exactly the resonantly localized field at the fundamental wavelength of 1550 nm stimulate the nonlinear response at the second-harmonic wavelength 775 nm, which have been observed in other nonlinear dielectric nanostructures. Accordingly, the multipolar contributions to the nonlinear response are also deduced from the displacement current distribution to elucidate the resonance behavior of the SHG efficiency. We can clearly see that the SHG efficiency is mainly dictated by the magnetic dipole contribution for the considered radius range. As a further confirmation, Figs. \ref{fig5}(b) and \ref{fig5}(c) display the generated nonlinear near-field distributions for the nanostructure with $r=200$ nm. The two electric field loops in the lateral direction with their constructive effect provide the fingerprint of the dominant magnetic dipole. 

\begin{figure}[htbp]
\centering
\includegraphics% Here is how to import EPS art
[scale=0.48]{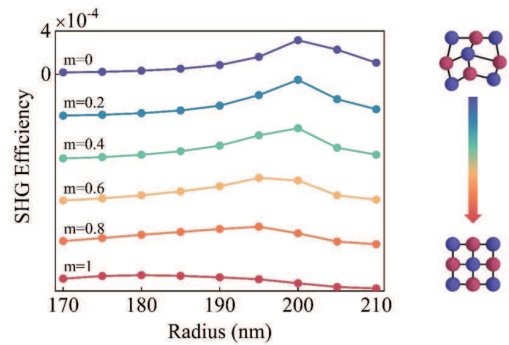}
\caption{\label{fig6} The SHG efficiency spectra as a function of the nanoantenna radius at the second-harmonic wavelength of 775 nm during the full process of GST amorphous-to-crystalline phase change. From the top to the bottom are different crystallization ratios, $m=0$ (a-GST), $m=0.2$, 0.4, 0.6, 0.8 (partial crystalline), and $m=1$ (c-GST), respectively.}
\end{figure}

Similar to the linear scattering, the nonlinear SHG process in the AlGaAs nanoantennas is to be tuned based on the precise control of GST phase change. In the framework, the tuning performance of the SHG efficiency is analyzed to obtain an insightful picture of the strong correlation between optical responses and the resonance supported by the dielectric nanostructures. Here we calculate the SHG efficiency $\eta_{\text{SHG}}$ of the proposed nanostructure at the second-harmonic wavelength of 775 nm when the GST film is in different crystallization ratios $m$, as shown in Fig. \ref{fig6}. $\eta_{\text{SHG}}$ goes through a multi-level modulation for each specific radius, and the maximum SHG efficiency shifts from $r=200$ nm with a-GST to $r=180$ nm with c-GST, exactly resembling the shift of the maximum linear scattering scattering in Fig. \ref{fig4}. The underlying physics as well originates from the modulation of resonant conditions by changes in optical properties of GST film. More specifically, the increase of $m$ leads to the larger real part $n$ of GST refractive index, and hence the AlGaAs nanoantennas have to shrink the sizes, i.e. the radius here, to satisfy the effective refractive index of the whole nanostructure required for maintaining the maximum linear scattering efficiency at the fixed fundamental wavelength of 1550 nm, and consequently rendering the maximum nonlinear SHG efficiency at the second-harmonic wavelength of 775 nm, which in turn validates the key role of the Mie-type resonances and the associated resonantly localized field in enhancing the SHG efficiency. This robust feature provide a helpful guide for smart designing tunable and reconfigurable nonlinear devices since the modulation of the nonlinear optical response can be accurately predicted with that of linear optical response in the same nanostructure. It is also interesting to notice the significant reduction in the maximum efficiency of the nonlinear SHG process from $3.15\times10^{-4}$ for $r=200$ nm with a-GST to $1.46\times10^{-4}$ for $r=180$ nm with c-GST, which is distinct from the tiny change in the maximum efficiency of the linear scattering. The difference, in fact, arises from the much larger absorption loss of the GST film around the second-harmonic wavelength of 775 nm than that around the fundamental wavelength of 1550 nm, due to a great increase in the imaginary part $k$ of the refractive index of GST. Such significant characteristic should be taken into full consideration in tuning the SHG efficiency as change in the GST film is reflected in strong perturbation on the SHG signal while it has little influence on the linear scattering. 

\begin{figure*}[htbp]
\centering
\includegraphics% Here is how to import EPS art
[scale=0.48]{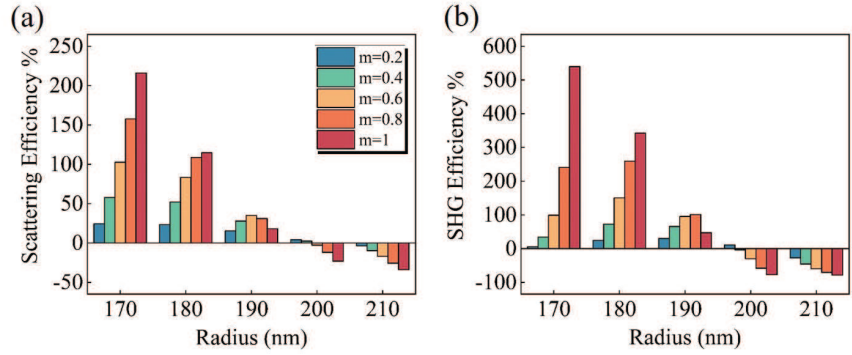}
\caption{\label{fig7} The modulation depths as a function of the nanoantenna radius (a) at the fundamental wavelength of 1550 nm and (b) at the second-harmonic wavelength of 775 nm during the full process of GST amorphous-to-crystalline phase change. From the left to the right are different crystallization ratios, $m=0$ (a-GST), $m=0.2$, 0.4, 0.6, 0.8 (partial crystalline), and $m=1$ (c-GST), respectively.}
\end{figure*}

To highlight the modulation of both the linear and nonlinear responses of the proposed nanostructure, we quantify the variation in the linear scattering efficiency and the nonlinear SHG efficiency, and provide the summarized results with different crystallization ratios. Here we introduce the normalized modulation depth,
\begin{equation}
\Delta\eta=\frac{\eta_{m}-\eta_{0}}{\eta_{0}},\label{eq5}
\end{equation}
where $\eta_{m}$ represents the linear scattering efficiency $\eta_{\text{SCAT}}$ or the nonlinear SHG efficiency $\eta_{\text{SHG}}$ with different crystallization ratios $m$, while $\eta_{0}$ is the corresponding efficiency with $m=0$ (a-GST). Under this definition, the results discussed in Figs. \ref{fig4} and \ref{fig6} are summarized, and the corresponding modulation depths are calculated in Figs. \ref{fig7}(a) and \ref{fig7}(b), respectively. Both the linear and nonlinear optical responses go through a multi-level modulation as GST crystallization ratio $m$ changes from 0 to 1, allowing to selectively tune the linear and nonlinear SHG signal via the precise control of GST phase change. This feature also reveals the robustness of the modulation method for the geometrical parameters in terms of the nanoantenna radius here. We can observe the positive $\Delta\eta$, i.e. an increase of both the scattering efficiency and nonlinear SHG efficiency for the radius smaller than 200 nm, as well as a continuous decrease of the efficiencies for the radius larger than 200 nm. These variation trends follow the shift of the Mie-type resonance at every step during the full process of GST amorphous-to-crystalline phase change. For the radius $r=200$ nm, an initial tiny increment of the efficiencies is surprisingly perceived (which is unconspicuous in Figs. \ref{fig4} and \ref{fig6}) and then the efficiencies turn into a decrease. This phenomenon suggests that the optimal AlGaAs nanoantenna radius which matches the fundamental wavelength for the maximum efficiencies may be not 200 nm but very close to 200 nm. Remarkably, the maximum value of $\eta_{\text{SHG}}$ in the SHG process reaches as high as $540\%$ for the radius $r=170$ with c-GST at the second-harmonic wavelength of 775 nm, as a result of the maximum $\eta_{\text{SCAT}}=216\%$ at the fundamental wavelength of 1550 nm.  

\begin{figure*}[htbp]
\centering
\includegraphics% Here is how to import EPS art
[scale=0.48]{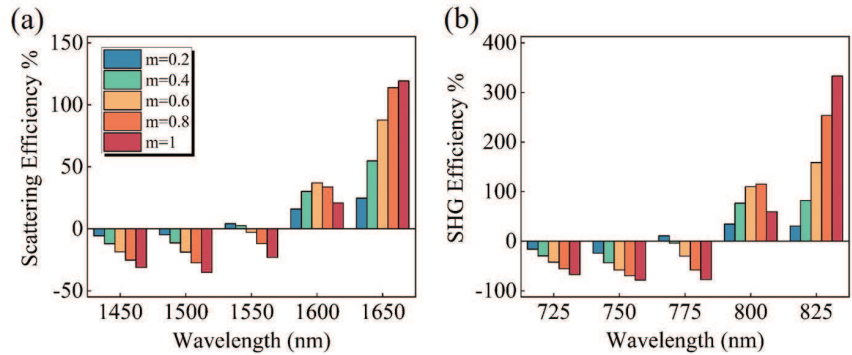}
\caption{\label{fig8} The modulation depths as a function of (a) the fundamental wavelength and (b) the second-harmonic wavelength for the radius $r=200$ nm during the full process of GST amorphous-to-crystalline phase change. From the left to the right are different crystallization ratios, $m=0$ (a-GST), $m=0.2$, 0.4, 0.6, 0.8 (partial crystalline), and $m=1$ (c-GST), respectively.}
\end{figure*}

In addition to the modulation at the fixed fundamental wavelength, we also explore the possibility of tuning the nonlinear optical response in a broad wavelength range. With the progressive change in the optical properties of GST, Figs. \ref{fig8}(a) and \ref{fig8}(b) provide the normalized modulation depth of linear and nonlinear efficiencies at different fundamental wavelengths for the fixed nanoantenna radius $r=200$ nm. As can be expected, the multi-level modulation of both the linear scattering and nonlinear SHG processes also applies to this case. The efficiencies change dramatically for different fundamental wavelengths, more specifically, the scattering efficiency and nonlinear SHG efficiency display continuous decrease with negative $\Delta\eta$ for the fundamental wavelength smaller than 1550 nm, and gradual increase with positive $\Delta\eta$ for the wavelength larger than 1550 nm. Keeping in mind that the increase in the effective refractive index of GST film from amorphous to crystalline phases brings about the larger resonant wavelength of the nanostructure, it is reasonable to conclude that for the fixed radius $r=200$ nm, the Mie-type resonance and the resulting maximum efficiency yield a redshift from the fundamental wavelength of 1550 nm to 1650 nm within the considered range. By virtue of the feature, it is desirable to dynamically tune the nonlinear response of the proposed nanostructure without altering the geometry. Note that, a large modulation depth of $333\%$ appears around second-harmonic wavelength of 825 nm with c-GST. The tuning strategy of SHG signal assisted by active material GST shows great flexibility and remarkable efficiency, offering additional degree of freedom to design the nonlinear nanostructures operating at different fundamental wavelengths.

\section{\label{sec4}Conclusions}

In conclusion, we propose a new solution for dynamically tunable SHG process via chalcogenide phase change material GST and demonstrate it in a typical nonlinear dielectric nanostructure at the telecommunication wavelength. In the proposed design consisting of the GST film deposited between the AlGaAs nanoantennas and AlO$_{x}$ substrate, the optically-induced Mie-type resonance enabled by electric and magnetic dipole modes provides the resonantly localized field for enhancing the nonlinear SHG efficiency, and the GST film plays a vital role in the modulation of resonant conditions. The numerical results show that the progressive changes in optical properties of GST lead to the gradual variation in the resonant properties of the AlGaAs nanoantennas in both the linear and nonlinear regions. The maximum modulation depth in the nonlinear SHG signal reaches as high as $540\%$ during the full process of GST amorphous-to-crystalline phase change, as a result of the maximum modulation depth of $216\%$ in the linear scattering efficiency. With the ultrafast phase change speed, high cyclability, and excellent thermal stability of GST, the proposed tuning strategy is expected to be rapidly tunable, reversible, multi-level, and nonvolatile, opening an avenue towards engineering fast, strong, and tunable optical nonlinearity. The results are the proof-of-principle demonstrations, and the modulation principle can be extended to higher harmonic generations and translated into the nonlinear metamaterials and metadevices if assembling the isolated nanoantennas into periodic arrays. Worth mentioning at this time is the remarkable third-order nonlinearity $\chi^{(3)}$ $\sim10^{-18}$ m$^{2}$/V$^{2}$ in chalcogenide glass as recently reported\cite{Cao2019, Yue2021}, which should be taken into account if this modulation method is used in the tunable and reconfigurable THG process.

\begin{acknowledgments}	
	
This work is supported by the National Natural Science Foundation of China (Grants No. 11847132, No. 11947065, No. 61901164, and No. 62005164), the Shanghai Rising-Star Program (Grant No. 20QA1404100), the Natural Science Foundation of Jiangxi Province (Grant No. 20202BAB211007), the Interdisciplinary Innovation Fund of Nanchang University (Grant No. 2019-9166-27060003), and the China Scholarship Council (Grant No. 202008420045).

\end{acknowledgments}

%\bibliography{Ref}% Produces the bibliography via BibTeX.


\end{document}